\documentclass[12pt]{article}

\usepackage{graphicx}
\usepackage{amsmath}
\usepackage{amssymb}
\usepackage{amsthm}
\usepackage{caption2}
\usepackage{setspace}
\usepackage[natural]{xcolor}

\newtheorem{theorem}{Theorem}[section]

\newtheorem{definition}{Definition}[section]

\DeclareMathOperator*{\doublesum}{\sum\sum}

\def\bi#1#2{ \ba{c} #1 \\ #2 \ea }
\def\mb{\mbox{ }}

\def\ba{\begin{array}}
\def\bd{\begin{document}}
\def\bdes{\begin{description}}
\def\bc{\begin{center}}
\def\be{\begin{equation}}
\def\bea{\begin{eqnarray}}
\def\beaa{\begin{eqnarray*}}
\def\bit{\begin{itemize}}
\def\ben{\begin{enumerate}}
\def\bt{\begin{tabular}}
\def\ea{\end{array}}
\def\ed{\end{document}}
\def\edes{\end{description}}
\def\ec{\end{center}}
\def\ee{\end{equation}}
\def\eea{\end{eqnarray}}
\def\eeaa{\end{eqnarray*}}
\def\et{\end{tabular}}
\def\eit{\end{itemize}}
\def\een{\end{enumerate}}
\def\ub{\underbrace}
\def\ob{\overbrace}
\def\ul{\underline}
\def\bv\begin{verbatim}
\def\ev\end{verbatim}
\def\bal{\begin{align*}}
\def\eal{\end{align*}}

\def\bfv#1{{\bf #1}}
\def\a{\alpha}
\def\b{\beta}
\def\g{\gamma}
\def\d{\delta}
\def\e{\varepsilon}
\def\z{\zeta}
\def\t{\theta}
\def\k{\kappa}
\def\p{\phi}
\def\o{\omega}
\def\G{\Gamma}
\def\D{\Delta}
\def\vp{\varphi}
\def\tdown{\bigtriangledown}
\def\diff{\nabla}
\def\T{\Theta}
\def\L{\Lambda}
\def\S{\Sigma}
\def\P{\Phi}
\def\O{\Omega}
\def\inf{\infty}
\def\prop{\propto}
\def\r|{\right|}
\def\l#1{\left#1}
\def\r#1{\right#1}
\def\ol{\overline}
\def\ltend{\longrightarrow}
\def\tend{\rightarrow}
\def\imply{\Longrightarrow}
\def\X{{\it X }}
\def\Y{{\it Y }}
\def\PB{\Phi (B)}
\def\TB{\Theta (B)}
\def\dd{(1-B)^d}
\def\PBinv{{\Phi (B)}^{-1}}
\def\TBinv{{\Theta (B)}^{-1}}
\def\SST{{\rm SST}}
\def\SSR{{\rm SSR}}
\def\SSE{{\rm SSE}}
\def\MSR{{\rm MSR}}
\def\MST{{\rm MST}}
\def\MSE{{\rm MSE}}
\def\mb{\mbox{ }}
\def\dv{ dependent variable }
\def\iv{ independent variable }
\def\mgf{ moment generating function }
\def\iid{ independent and identically distributed }
\def\pdf{ probability density function }
\def\rv{ random variable }
\def\rvs{ random variables }
\def\varx{{\sigma_x}^2}
\def\vary{{\sigma_y}^2}
\def\rhoxy{{\rho_{x,y}}}
\def\covxy{{Cov( X , Y )}}
\def\bra#1{{( #1_i - \bar #1 )}}
\def\bras#1{{( #1_i - \bar #1 )}^2}
\def\oon{\frac 1{n-1}}
\def\vis#1{#1_i^2}
\def\vbs#1{\bar #1^2}
\def\vsvh#1#2{{\hat #1}_{#2}}
\def\vhi#1{\hat #1_i}
\def\vih#1{\hat #1_i}
\def\bxy#1{( x_#1 , y_#1 )}
\def\tnt{t_{(n-2)}}
\def\tnta{t_{(n-2,\alpha)}}
\def\tntas{t_{(n-2,\alpha /2)}}
\def\veps{\varepsilon}
\def\vepsi{\veps_i}
\def\vepsj{\veps_j}
\def\vareps{\sigma_\veps^2}
\def\var{\sigma^2}
\def\svs#1{s_#1^2}
\def\yih{{\hat y}_i}
\def\binom#1#2{\left(\ba{c} #1 \\ #2 \ea \right)}
\def\bi#1#2{ \ba{c} #1 \\ #2 \ea }
\def\s#1#2{#1_#2}
\def\sxx#1#2#3{#1_{#2 #3}}
\def\sxxx#1#2#3#4{#1_{#2#3#4}}
\def\sxxxx#1#2#3#4#5{#1_{#2#3#4#5}}
\def\sxxxxx#1#2#3#4#5#6{#1_{#2#3#4#5#6}}
\def\sxxxxxx#1#2#3#4#5#6#7{#1_{#2#3#4#5#6#7}}
\def\sxxxxxxx#1#2#3#4#5#6#7#8{#1_{#2#3#4#5#6#7#8}}
\def\h#1{\hat #1}
\def\sh#1#2{{\hat #1}_#2}
\def\sxxh#1#2#3{{\hat #1}_{#2 #3}}
\def\sxxxh#1#2#3#4{{\hat #1}_{#2#3#4}}
\def\sxxxxh#1#2#3#4#5{{\hat #1}_{#2#3#4#5}}
\def\sxxxxxh#1#2#3#4#5#6{{\hat #1}_{#2#3#4#5#6}}
\def\sxxxxxxh#1#2#3#4#5#6#7{{\hat #1}_{#2#3#4#5#6#7}}
\def\sxxxxxxxh#1#2#3#4#5#6#7#8{{\hat #1}_{#2#3#4#5#6#7#8}}
\def\ss#1#2#3{#1_{#2,#3}}
\def\ts#1#2#3{#1_{#2_#3}}
\def\tsxx#1#2#3#4#5{#1_{{#2_#3}#4#5}}
\def\sumv#1#2#3{\sum_{#1=#2}^{{#3}}}
\def\chis{\chi^2}
\def\vsp#1#2#3{#1_#2^#3}
\def\sfss#1#2{S_{#1 + #2}^f}
\def\SS#1#2{{\mbox{SS}}_{#1 #2}}
\def\mb{\mbox{ }}
\def\n0v{N( 0 ,\sigma^2 )}
\def\nmv{N( \mu ,\sigma^2 )}
\def\bvsv#1#2{\{ #1_#2 \}}

\begin{document}

\thispagestyle{empty}

\baselineskip = 18pt
\begin{center}
{\bf \Large {\bf The Hiemstra-Jones Test Revisited}}  \large
\\
\vskip 1ex
{\bf Zhidong Bai} \\ School of Mathematics and Statistics, Northeast Normal University \\
\vskip 1ex {\bf Yongchang Hui}\footnote{Yongchang Hui, School of Mathematics and Statistics,
Xi'an Jiaotong University. No.28, Xianning West Road, Xi'an, Shaanxi, P.R. China.
Tel: (86)-029-82663170, Email: huiyc180@xjtu.edu.cn .} \\ School of Mathematics and Statistics, Xi'an Jiaotong University \\
\vskip 1ex {\bf Zhihui Lv} \\School of Mathematics and Statistics, Northeast Normal University \\
\vskip 1ex {\bf Wing-Keung Wong} \\ Department of Finance, Asia University, Taiwan \\
Department of Economics, Lingnan University, Hong Kong \\
\vskip 1ex {\bf Zhen-Zhen Zhu} \\School of Mathematics and Statistics, Northeast Normal University \\
\end{center}



\vspace{0.1in} \noindent \bf Abstract \rm $\quad$
The famous Hiemstra-Jones (HJ) test developed by Hiemstra and Jones (1994)
plays a significant role in studying nonlinear causality. Over the last two decades, there have been numerous applications and theoretical extensions based on this pioneering work. However, several works note that counterintuitive results are obtained from the HJ test, and some researchers find that the HJ test is seriously over-rejecting in simulation studies. In this paper, we reinvestigate HJ's creative 1994 work and find that their proposed estimators of the probabilities over different time intervals were not consistent with the target ones proposed in their criterion. To test HJ's novel hypothesis on Granger causality, we  propose new estimators of the probabilities defined in their paper and reestablish the asymptotic properties to induce new tests similar to those of HJ. Some simulations will also be presented to support our findings.

\vspace{0.1in} \noindent \bf Keywords: \rm $\quad$  Central limit theorem, Hiemstra-Jones test, Nonlinear Granger causality.

\newpage
\baselineskip = 20pt

\begin{center}
{\bf \Large {\bf Conflict-of-interest disclosure statement}}

\vspace{0.2in}

\begin{enumerate}
  \item {\textbf{Zhidong Bai}. I declare that there is no conflict of interest.}
  \item {\textbf{Yongchang Hui}. I declare that there is no conflict of interest.}
  \item {\textbf{Zhihui Lv}. I declare that there is no conflict of interest.}
  \item {\textbf{Wing-Keung Wong}. I declare that there is no conflict of interest.}
  \item {\textbf{Zhen-Zhen Zhu}. I declare that there is no conflict of interest.}
\end{enumerate}
\end{center}


\thispagestyle{empty}

\newpage \setcounter{page}{1}
\section{Introduction} \label{Introduction}
After the pioneering work of Granger (1969), Granger causality tests have developed into a set of useful methods to detect causal relations between time series in economics and finance.
Consider a strictly stationary bivariate time series $\{(X_t, Y_t)\}$, $t
\in Z$. Intuitively, $\{Y_t\}$ is a
Granger cause of $\{X_t\}$ if adding past observations of $Y_t$ to the
information set increases knowledge about the distribution of
current values of $X_t$.

Linear Granger causality tests within the linear autoregressive
model class have been developed in many directions, e.g., Hurlin et al (2001) proposed
a procedure for causality tests with panel data, Bai et al (2008)
extend the traditional bivariate Granger causality test to multivariate situations, Ghysels et al (2016)
test for Granger causality with mixed frequency data based on the multiple-horizon framework
established by Dufour and Renault (1998) and Dufour et al (2006).

Though linear tests of Granger causality have been investigated very deeply, they are limited in their capability to
detect nonlinear causality. 
There is no need emphasize the importance of nonlinear structures between variables, since the real world is ``almost certainly nonlinear,'' as Granger (1989) notes.
Modern developments in computer science and computing facilities motivate
ever increasing interest in testing nonlinear Granger causality. Among the various tests of nonlinear
Granger causality, the Hiemstra and Jones (1994) test (hereafter, the HJ test) 
is the most cited by scholars and the most frequently applied by practitioners in economics and finance.
There were over 1100 Google Scholar hits by September 2016, which illustrates its significance in the economics and finance literatures. However, some doubts about the efficacy of the HJ test arise from the many counterintuitive results.

Diks and Panchenko (2005) find two serious problems with the HJ test.
First, even if there is a strong evidence of linear Granger causality, the HJ test
can fail to detect causality. Second, using simulation studies, they show that under the null hypothesis,
the reject rate tends to 1 when sample size increases.

In accordance with the evidence presented by Diks and Panchenko (2005, 2006), in this paper, we reinvestigate the HJ test (1994) and reveal
some of the underlying reasons for this questionable performance. The remainder of this paper is organized as follows. In Section 2, we simply review the procedure of the HJ test.
In Section 3, we describe the crux of the problem identified by Diks and Panchenko (2005) and revise it accordingly.
Specifically, we re-estimate the probabilities defined by Hiemstra and Jones (1994) and deduce the asymptotic distribution of the test statistics.
Simulation results are presented in Section 4. Finally, we provide some concluding remarks in Section 5.

\vspace{.2in}

\section{Hiemstra-Jones Nonlinear Causality Test}\label{BivariateNLCT}
\bigskip

\bigskip
Hiemstra and Jones (1994) consider a causality test between two
strictly stationary and weakly dependent time series processes $\{X_t\}$
and $\{Y_t\}$. The $m$-length lead vector of $X_t$,
${L_x}$-length lag vector of $X_{t}$ and ${L_y}$-length lag vector of $Y_{t}$ are defined as
\begin{eqnarray*}
X_t^{m} &\equiv& \big (X_{t},X_{t+1}, \cdots,X_{t+m-1}\big ),
~m=1,2,\cdots,~t=1,2,\cdots \\
X_{t-{L_x}}^{{L_x}} &\equiv& \big (X_{t-{L_x}},X_{t-{L_x}} ,
\cdots,X_{t-1} \big ), {L_x}=1,2,\cdots, t={L_x}+1, {L_x}+2,\cdots \\
Y_{t-{L_y}}^{{L_y}} &\equiv& \big (Y_{t-{L_y}},Y_{t-{L_y}} ,
\cdots,Y_{t-1} \big ), {L_y}=1,2,\cdots, t={L_y}+1, {L_y}+2,\cdots
\, .
\end{eqnarray*}

Hiemstra and Jones (1994) 
define non-Granger causality from $\{Y_t\}$ to $\{X_t\}$.
\begin{definition}  \label{defsinglenonlinear} \quad
For any given values of $m$, $L_x$, $L_y > 1$ and $e > 0$, series $\{Y_t\}$ does not strictly
Granger cause $\{X_t\}$ if
\begin{eqnarray} \label{def1}
P \left (\|X_t^m-X_s^m\|<e | \parallel
X_{t-{L_x}}^{{L_x}}-X_{s-{L_x}}^{{L_x}}\parallel<e,\parallel
Y_{t-L_y}^{L_y}-Y_{s-L_y}^{L_y}\parallel<e  \right )
\nonumber \\
= P \left (\|X_t^m-X_s^m\|<e  | \parallel
X_{t-{L_x}}^{{L_x}}-X_{s-{L_x}}^{{L_x}}\parallel<e  \right )
\, ,
\end{eqnarray}
where $Pr( \cdot \, | \, \cdot \, )$ denotes the conditional probability
and $\parallel \cdot \parallel$ denotes the maximum norm, which is
defined as $\|X-Y\|= max  \big ( |x_1-y_1|, |x_2-y_2|, \cdots,
|x_n-y_n| \big )$ for any two vectors $X= \big (x_1, \cdots, x_n
\big )$ and $Y= \big (y_1, \cdots, y_n \big )$.
\end{definition}

Using the notation
\begin{align*}
& C_1 \big({m}+{L_x},{L_y},e \big)\equiv Pr \left  (\parallel
X_{t-{L_x}}^{{m}+{L_x}}-X_{s-{L_x}}^{{m}+{L_x}}\parallel
<e,\parallel
Y_{t-{L_y}}^{{L_y}}-Y_{s-{L_y}}^{{L_y}}\parallel <e \right ) \, , \\
& C_2 \big({L_x},{L_y},e \big)\equiv Pr \left (\parallel
X_{t-{L_x}}^{{L_x}}-X_{s-{L_x}}^{{L_x}}\parallel<e,\parallel
Y_{t-{L_y}}^{{L_y}}-Y_{s-{L_y}}^{{L_y}}\parallel<e \right ) \, , \,\\
& C_3 \big({m}+{L_x},e \big)\equiv Pr \left (\parallel
X_{t-{L_x}}^{{m}+{L_x}}-X_{s-{L_x}}^{{m}+{L_x}}\parallel <e
\right ) \, ,
\, \, \mbox{and} \\
& C_4 \big({L_x},e \big)\equiv Pr \left (\parallel
X_{t-{L_x}}^{{L_x}}-X_{s-{L_x}}^{{L_x}}\parallel<e \right ) \, ,
\end{align*}
Hiemstra and Jones (1994) re-express Equation (\ref{def1}) as
\begin{equation}
\frac{C_1 \big ({m}+{L_x},{L_y},e \big )}{C_2 \big
({L_x},{L_y},e \big )} = \frac{C_3 \big ({m}+{L_x},e \big )}{C_4
\big ({L_x},e  \big)} \, .
\end{equation}

After this preparation, they propose the following nonlinear Granger
causality test statistic
\begin{equation}
\sqrt{n}\left(\frac{C_1 \big (m+{L_x},L_y,e,n \big )}{C_2 \big
({L_x},L_y,e,n
 \big )}-\frac{C_3 \big (m+{L_x},e,n \big )}{C_4 \big ({L_x},e,n \big )}\right)\, , \label{test1}
\end{equation}
where
\begin{eqnarray*}
&& C_1  \big (m+{L_x},L_y,e,n  \big ) \equiv
\frac{2}{n(n-1)}\doublesum_{t<s} I  \big
(x_{t-{L_x}}^{m+{L_x}},x_{s-{L_x}}^{m+{L_x}},e  \big )
\cdot I \left (y_{t-L_y}^{L_y},y_{s-L_y}^{L_y},e \right ) \, , \\
&& C_2  \big ({L_x},L_y,e,n  \big ) \equiv
\frac{2}{n(n-1)}\doublesum_{t<s} I \left
(x_{t-{L_x}}^{{L_x}},x_{s-{L_x}}^{{L_x}},e \right )
\cdot I \left (y_{t-L_y}^{L_y},y_{s-L_y}^{L_y},e\right ) \, ,  \\
&& C_3 \big (m+{L_x},e,n \big )\equiv
\frac{2}{n(n-1)}\doublesum_{t<s}
I \left (x_{t-{L_x}}^{m+{L_x}}, x_{s-{L_x}}^{m+{L_x}},e \right )\, ,  \\
&& C_4 \big ({L_x},e,n \big )\equiv \frac{2}{n(n-1)}\doublesum_{t<s}
I \left (x_{t-{L_x}}^{{L_x}},x_{s-{L_x}}^{{L_x}},e \right ) \, , \,
\, \mbox{and} \\ && I(x,y,e)=
\begin{cases} 0, & \text{if $\|x-y\|>e$}\\
1, & \text{if $\|x-y\|\leq e$}
\end{cases} \, .
\end{eqnarray*}

They claimed that the $C_j(*,n)$s were $U$-statistic estimators of their counterparts $C_j(*)$ and tried to show the limiting results for the test statistics (\ref{test1}).
Although the estimators $C_j(*,n)$ looked like $U$-statistics, they were not because the expectations of the general terms are not the same. Moreover, the
$C_j(*)$s are related to the indices $t$ and $s$ (in fact, to $|t-s|$ for strongly stationary processes). The $C_j(*,n)$s were independent of $t$ and $s$ for summing up over them. 
 Therefore, the $C_j(*,n)$ estimators are neither consistent nor asymptotic normal estimators of their counterparts $C_j(*)$. Based on this analysis, one sees that the center of statistic (\ref{test1}) should tend toward infinity; hence, the test must be over-rejecting when the sample size is large.



\section{A new test of Hiemstra-Jones Nonlinear Causality}

It is worth reminding the reader that the pair $(s,t)$ (in fact, $|t-s|$ for strongly stationary processes) in Equation (\ref{def1})
of Definition \ref{defsinglenonlinear} is a key parameter of the probabilities $C_j(*)$.
In fact, Hiemstra and Jones (1994) note this, and there is no problem in Equation (\ref{def1})
of Definition \ref{defsinglenonlinear}. However, it seems that Hiemstra and Jones (1994) overlooked this fact in their proposed estimation of $C_j(*)$.
The improper estimators $C_j(*, n)$ thus lead to an invalid asymptotic distribution of the test statistic.

We now begin to state the procedure for our new test. For any given pair $(s,t)$, we denote
\begin{align*}
& C_1 \big({m}+{L_x},{L_y},e;t,s \big)\equiv Pr \left  (\parallel
X_{t-{L_x}}^{{m}+{L_x}}-X_{s-{L_x}}^{{m}+{L_x}}\parallel
<e,\parallel
Y_{t-{L_y}}^{{L_y}}-Y_{s-{L_y}}^{{L_y}}\parallel <e \right ) \, , \\
& C_2 \big({L_x},{L_y},e;t,s \big)\equiv Pr \left (\parallel
X_{t-{L_x}}^{{L_x}}-X_{s-{L_x}}^{{L_x}}\parallel<e,\parallel
Y_{t-{L_y}}^{{L_y}}-Y_{s-{L_y}}^{{L_y}}\parallel<e \right ) \, , \,\\
& C_3 \big({m}+{L_x},e;t,s \big)\equiv Pr \left (\parallel
X_{t-{L_x}}^{{m}+{L_x}}-X_{s-{L_x}}^{{m}+{L_x}}\parallel <e
\right ) \, ,
\, \, \mbox{and} \\
& C_4 \big({L_x},e;t,s \big)\equiv Pr \left (\parallel
X_{t-{L_x}}^{{L_x}}-X_{s-{L_x}}^{{L_x}}\parallel<e \right ) \, .
\end{align*}
Furthermore, we have
\begin{align*}
&Pr \left (\|X_t^{{m}}-X_s^{{m}}\|<e \big | \parallel
X_{t-{L_x}}^{{L_x}}-X_{s-{L_x}}^{{L_x}}\parallel<e,\parallel
Y_{t-{L_y}}^{{L_y}}-Y_{s-{L_y}}^{{L_y}}\parallel<e \right )   \\
& \quad =\frac{C_1 \big({m}+{L_x},{L_y},e;t,s \big)}{C_2 \big({L_x},{L_y},e;t,s \big)} \, , \,
\end{align*}
and
\begin{align*}
&Pr \left (\|X_t^{{m}}-X_s^{{m}}\|<e \big | \parallel
X_{t-{L_x}}^{{L_x}}-X_{s-{L_x}}^{{L_x}}\parallel<e \right )   \\
& \quad =\frac{C_3 \big({m}+{L_x},e;t,s \big)}{C_4 \big({L_x},e;t,s \big)} \, .
\end{align*}

\bigskip

Under the assumption of stationarity, for the given $(t,s)$ with $s-t=l$, we express
\begin{align*}
& C_1 \big({m}+{L_x},{L_y},e;t,s \big) \equiv C_1 \big({m}+{L_x},{L_y},e;t,l \big) = C_1 \big({m}+{L_x},{L_y},e;l \big) \, , \\
& C_2 \big({L_x},{L_y},e;t,s \big)\equiv C_2 \big({L_x},{L_y},e;t,l \big) = C_2 \big({L_x},{L_y},e;l \big)\, , \,\\
& C_3 \big({m}+{L_x},e;t,s \big)\equiv C_3 \big({m}+{L_x},e;t,l \big) = C_3 \big({m}+{L_x},e;l \big) \, ,
\, \, \mbox{and} \\
& C_4 \big({L_x},e;t,s \big)\equiv C_4 \big({L_x},e;t,l \big) = C_4 \big({L_x},e;l \big) \, .
\end{align*}

Thus, $\{Y_t\}$ does not strictly Granger cause another series $\{X_t\}$ nonlinearly, which means that  for each $l > 0$,
${C_1 \big ({m}+{L_x},{L_y},e;l \big )}/{C_2 \big
({L_x},{L_y},e;l \big )} = {C_3 \big ({m}+{L_x},e;l \big )}/{C_4
\big ({L_x},e;l  \big)}$.

If we now consider two sets of samples $\{x_1, x_2, \cdots  ,x_T\}$ and $\{y_1, y_2, \cdots  ,y_T\}$, we can examine whether there is
nonlinear Granger causality from $\{Y_t\}$ to $\{X_t\}$. That is, we  test the following hypothesis
\begin{equation}
H_{0} \  : \ \ \frac{C_1 \big ({m}+{L_x},{L_y},e;l \big )}{C_2 \big
({L_x},{L_y},e;l \big )} = \frac{C_3 \big ({m}+{L_x},e;l \big )}{C_4
\big ({L_x},e;l  \big)} \, .
\label{newh0}
\end{equation}

We first provide the consistent estimators of
 ${C_1 \big ({m}+{L_x},{L_y},e;l \big )}$, ${C_2 \big({L_x},{L_y},e;l \big )}$, ${C_3 \big ({m}+{L_x},e;l \big )}$ and ${C_4
\big ({L_x},e;l \big)}$
\begin{eqnarray*}
&& \hat{C}_1  \big (m+{L_x},L_y,e;l \big ) \equiv
\frac{1}{n}\sum\limits_{t=L_{xy}+1}^{T-l-m+1} I  \big
(x_{t-{L_x}}^{m+{L_x}},x_{t+l-{L_x}}^{m+{L_x}},e  \big )
\cdot I \left (y_{t-L_y}^{L_y},y_{t+l-L_y}^{L_y},e \right ) \, , \\
&& \hat{C}_2  \big ({L_x},L_y,e;l \big ) \equiv
\frac{1}{n}\sum\limits_{t=L_{xy}+1}^{T-l-m+1} I \left
(x_{t-{L_x}}^{{L_x}},x_{t+l-{L_x}}^{{L_x}},e \right )
\cdot I \left (y_{t-L_y}^{L_y},y_{t+l-L_y}^{L_y},e\right ) \, ,  \\
\end{eqnarray*}
\begin{eqnarray*}
&& \hat{C}_3 \big (m+{L_x},e; l \big )\equiv
\frac{1}{n}\sum\limits_{t=L_{xy}+1}^{T-l-m+1}
I \left (x_{t-{L_x}}^{m+{L_x}}, x_{t+l-{L_x}}^{m+{L_x}},e \right )\, ,  \\
&& \hat{C}_4 \big ({L_x},e;l \big )\equiv \frac{1}{n}\sum\limits_{t=L_{xy}+1}^{T-l-m+1}
I \left (x_{t-{L_x}}^{{L_x}},x_{t+l-{L_x}}^{{L_x}},e \right ) \, , \,
\, \\ && \mbox{where} \ L_{xy} = \max({L_x},{L_y}), \ I(x,y,e) =
\begin{cases} 0, & \text{if $\|x-y\|>e$}\\
1, & \text{if $\|x-y\|\leq e$}
\end{cases} \,  \\  && \mbox{and} \ n=T-L_{xy}-l-m+1.
\end{eqnarray*}

The consistency of our proposed estimators can be shown straightforwardly and is omitted from this paper.
We use a simple numerical study to show that our estimators are consistent whereas those of HJ are not.
Let $X_t = 2\varepsilon_{t-1} + \varepsilon_{t}$, $\varepsilon_{t} \stackrel{iid}{\sim} N(0,1)$, while \{$Y_t$\} could be any stationary sequence. Let $l = 1$, $L_x = L_y = m = 1$. We can calculate the exact values of ${C_4 \big ({L_x},e;l  \big)}$, which are $0.3169$ and $0.4597$, respectively, when $e= 1$ and $e= 1.5$. For simplicity, we denote the values of ${C_4 \big ({L_x},e;l  \big)}$ as $C_4$ and the HJ estimate and our estimate $\hat{C}_{4}^{HJ}$ and $\hat{C}_4$, respectively, in Table 1. Additionally, Table 1 provides the estimated values with their corresponding relative estimation errors in brackets when $T= 1000, 2000 \, \textrm{and} \, 4000$. It is obvious that the HJ estimator is not  consistent.

\begin{table}[!htbp]\scriptsize
\begin{center}
\caption{${C_4 \big ({L_x},e;l  \big)}$ and its estimated values.}  \label{table5} \vskip 0.1in
\begin{tabular}{|c|ccccccccc|}
\hline \multicolumn{1}{|c|}{} &\multicolumn{1}{c}{}
&\multicolumn{3}{c}{$e=1$} &\multicolumn{1}{c}{}
&\multicolumn{3}{c}{$e=1.5$} &\multicolumn{1}{c|}{}\\
\cline{3-5}\cline{7-9}
$T =$ && $C_4$ &$\hat{C}_4$&$\hat{C}_{4}^{HJ}$&&$C_4$&$\hat{C}_4$& $\hat{C}_{4}^{HJ}$& \\
\hline
1000&&0.3169&0.3056(3.56\%)&0.2564(19.0\%)&&0.4597 &0.4529(1.46\%) & 0.3755(18.2\%)& \\
\hline
2000&&0.3169 &0.3109(1.89\%)&0.2531(20.1\%)&&0.4597 & 0.4629(0.69\%)& 0.3718(19.1\%)&\\
\hline
4000 &&0.3169  &0.3128(1.29\%)&0.2474(21.9\%)&&0.4597 & 0.4599(0.05\%)&0.3636(20.9\%)&\\
\hline
\end{tabular}\\
\begin{flushleft}
{\small Note:
The true value of ${C_4 \big ({L_x},e;l  \big)}$ is denoted $C_4$, the HJ estimate and our estimate are denoted $\hat{C}_{4}^{HJ}$ and $\hat{C}_4$, respectively. The relative estimation errors are in the accompanying brackets.}
\end{flushleft}
\end{center}
\end{table}

\bigskip
Now, we propose
\begin{equation}
T_n = \sqrt{n}\left(\frac{\hat{C}_1 \big (m+{L_x},L_y,e,l \big )}{\hat{C}_2 \big
({L_x},L_y,e,l \big )}-\frac{\hat{C}_3 \big (m+{L_x},e,l \big )}{\hat{C}_4 \big
({L_x},e,l \big )}\right)
\end{equation}
as the test statistic, and we establish the following asymptotic distribution of $T_n$ for statistical inference.

\begin{theorem} \label{BHWtest} \quad Stationary sequences $\{x_t, t=1, \cdots, T\}$ and $\{y_t, t=1, \cdots, T\}$
are both strong mixing, with mixing coefficients satisfying the conditions of Lemma 1 presented in Appendix, for given values of
$l, L_x, L_y, m$ and $e>0$, under the null hypothesis that $\{y_t\}$ does not strictly
Granger cause $\{x_t\}$, then the test statistic is defined in
(\ref{test1})
\begin{equation*}
\sqrt{n}\left(\frac{\hat{C}_1 \big (m+{L_x},L_y,e,l \big )}{\hat{C}_2 \big
({L_x},L_y,e,l \big )}-\frac{\hat{C}_3 \big (m+{L_x},e,l \big )}{\hat{C}_4 \big
({L_x},e,l \big )}\right) \overset{d}{\longrightarrow} N \big(0, \sigma^2
(m,{L_x},L_y,e,l) \big) \, .
\end{equation*}
\end{theorem}

The asymptotic variance $\sigma^2(m,{L_x},L_y,e,l)$ with its consistent
estimator $\hat{\sigma}^2(m,{L_x},L_y,e,l)$ and the proof of theorem \ref{BHWtest} are given in the Appendix. The hypothesis $H_0$ defined in (\ref{newh0}) is rejected at $\alpha$ if
\begin{eqnarray*}
\big| T_n \big |/{\hat{\sigma}^2(m,{L_x},L_y,e,l)} > z_{{\alpha/2}},
\end{eqnarray*}
where $z_{{\alpha/2}}$ is the up ${\alpha/2}$ quantile of the standard normal distribution.
In this situation, we will conclude that there exists nonlinear Granger causality from \{$Y_{t}$\} to \{$X_{t}$\}.

There are several possible methods to estimate the asymptotic covariance $\sigma^2(m,{L_x},L_y,e,l)$. A model-based approach uses known laws of \{$X_t$\} and \{$Y_t$\} to calculate the expectations in the formula given in the Appendix and simply substitutes $C_j(*), j = 1, 2, 3,4$ with their corresponding estimates. However, in practice, we can hardly avoid model misspecification and may obtain improper laws of \{$X_t$\} and \{$Y_t$\}. We suggest the use of bootstrap methods as in the simulation studies we use to test hypothesis $H_0$.

\vspace{.2in}
\section{Simulation}
In this section, we perform numerical studies using simulations to illustrate the applicability
and superiority of the new nonlinear Granger causality test developed in Section 3.
Let $R$ be the times of rejecting the null hypothesis that $\{Y_t\}$ does not strictly Granger cause $\{X_t\}$ nonlinearly in 10,000 replications at the
$\alpha$ level, and thus, the empirical power is ${R}/{10,000}$. In our simulation, the length of the testing sequences is 1000, and we chose the same lag length and lead length: $L_x = L_y = m = 1$. We set three situations of $l$ and two situations of $e$: $l=1$, $l=2$, $l=3$ and $e=1$, $e=1.5$.
Consider the following two cases.

\begin{itemize}
\item Case 1: $(X_t,Y_{t-1}) \stackrel{iid}{\sim} N(0,\Sigma)$,
$\Sigma = \left(\begin{array}{cc}
  1 & \rho \\
  \rho & 1
\end{array}\right)$. $\rho = 0, 0.2, \cdots, 0.8$.
\item Case 2: $X_t = \sqrt{1+0.4X_{t-1}^{2}}\varepsilon_{t}$, $Y_{t-1} = {\rho}X_{t}+\eta_{t}$, \ where \\ \ \ \ \ \ \ \ \ \ \ $\varepsilon_{t} \bot \eta_{t}$, $\varepsilon_t \stackrel{iid}{\sim} N(0,1)$, $\eta_t \stackrel{iid}{\sim} N(0,1)$. $\rho = 0, 0.2, \cdots, 0.8$.
\end{itemize}

\begin{figure}[!htbp]
\caption{Test nonlinear Granger causality form $Y_t$ to $X_t$: Case 1}
\begin{center} \
\includegraphics[height=6cm,width=4cm]{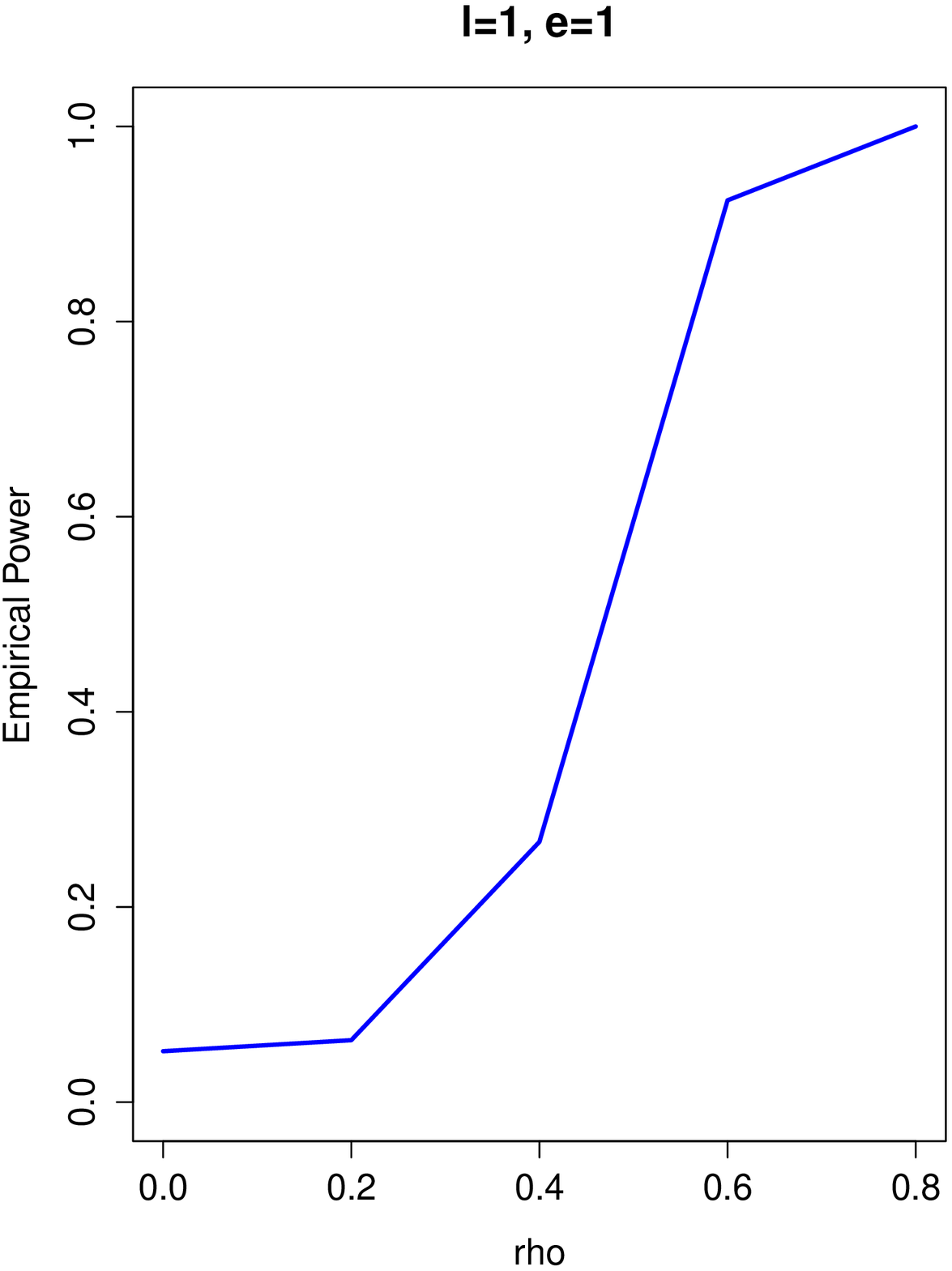}
\includegraphics[height=6cm,width=4cm]{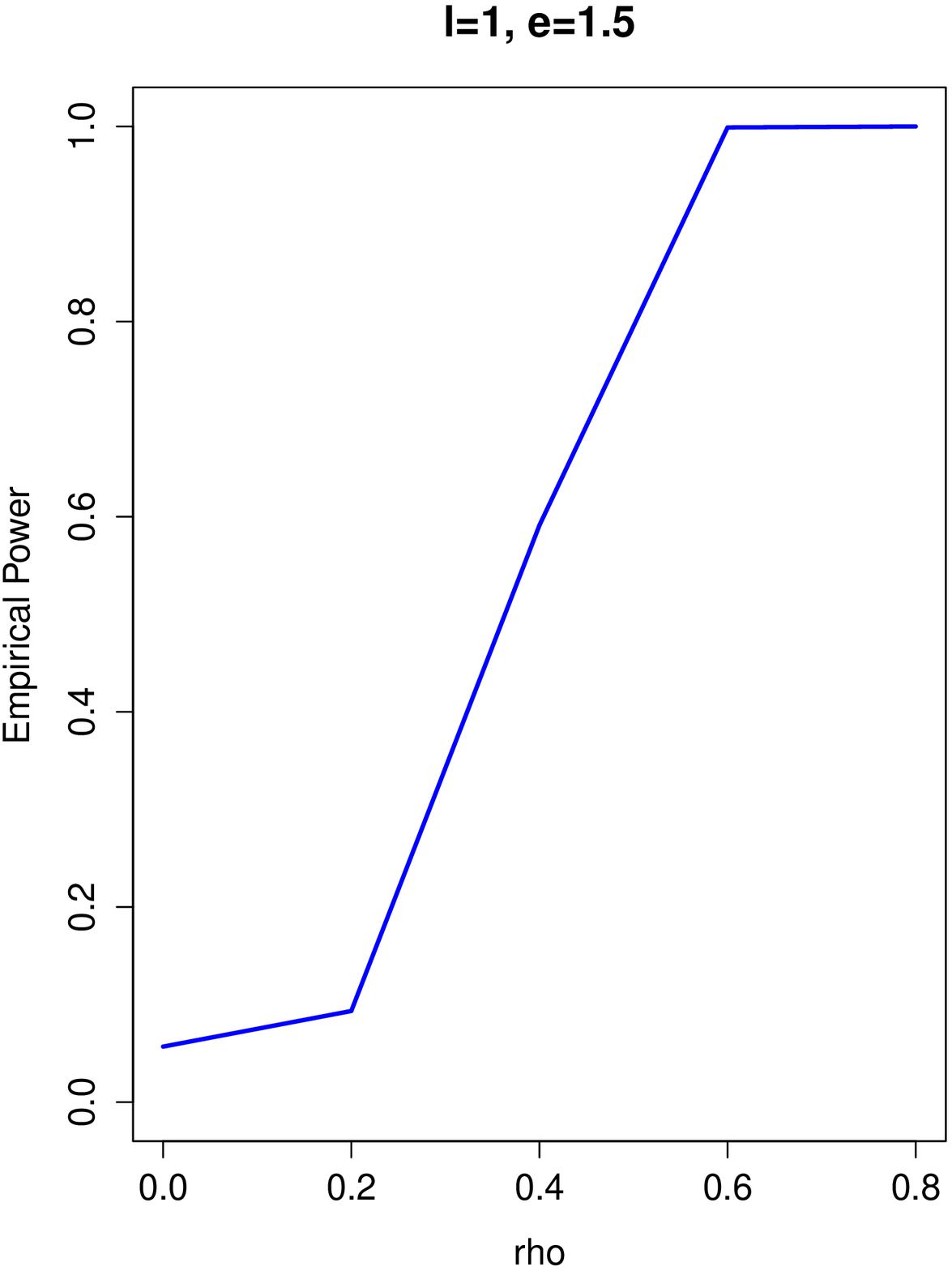} \\
\includegraphics[height=6cm,width=4cm]{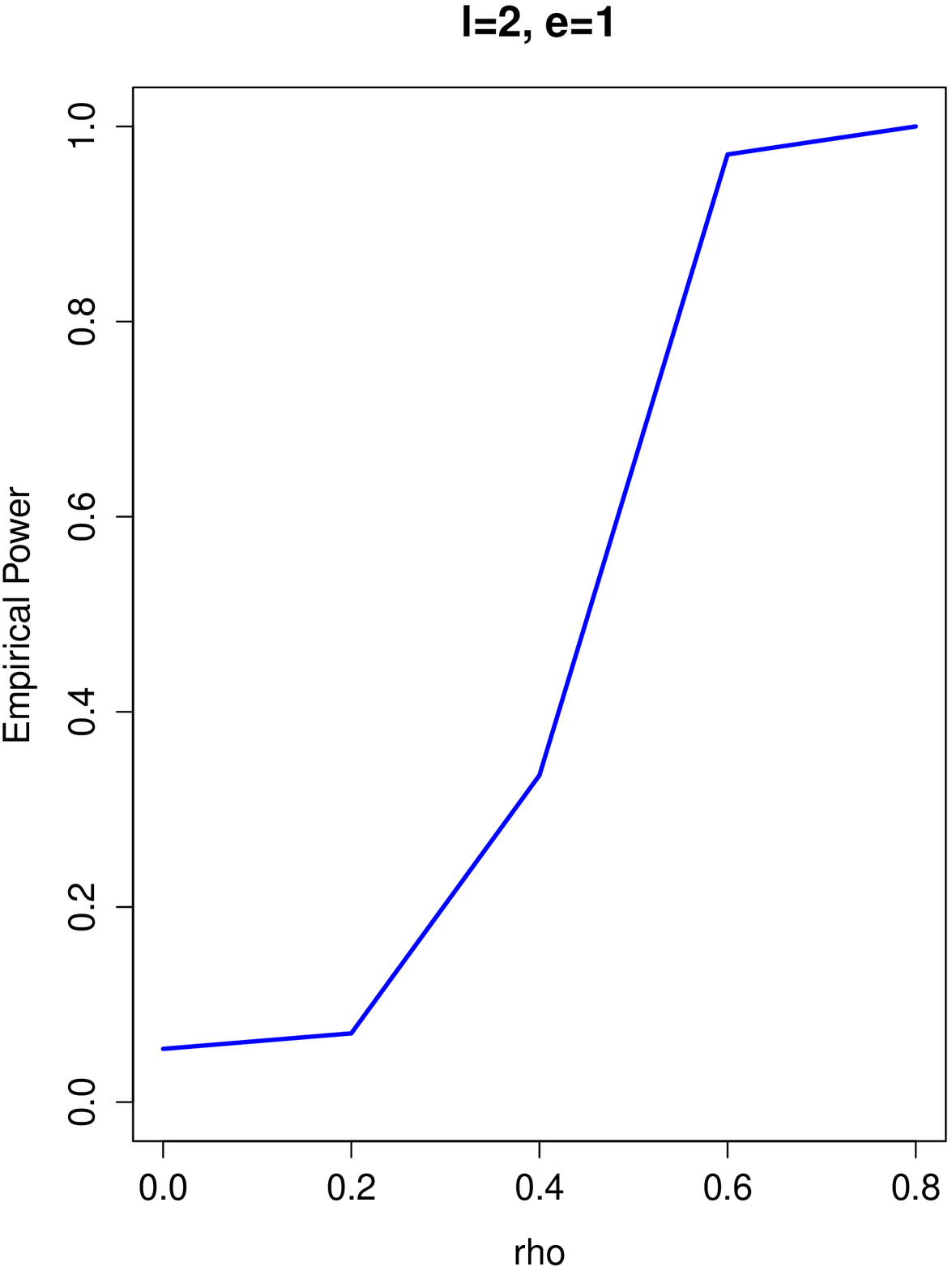}
\includegraphics[height=6cm,width=4cm]{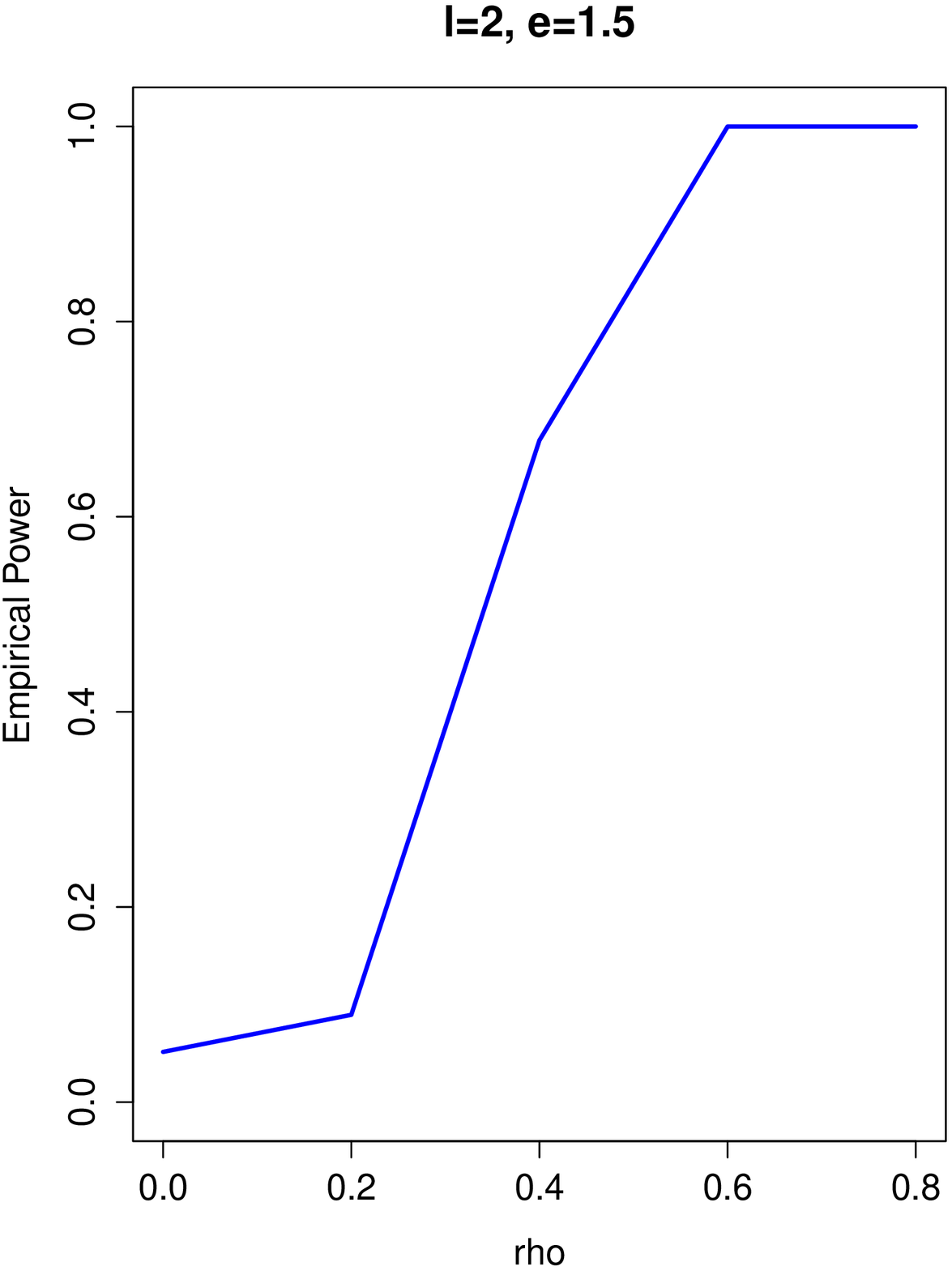} \\
\includegraphics[height=6cm,width=4cm]{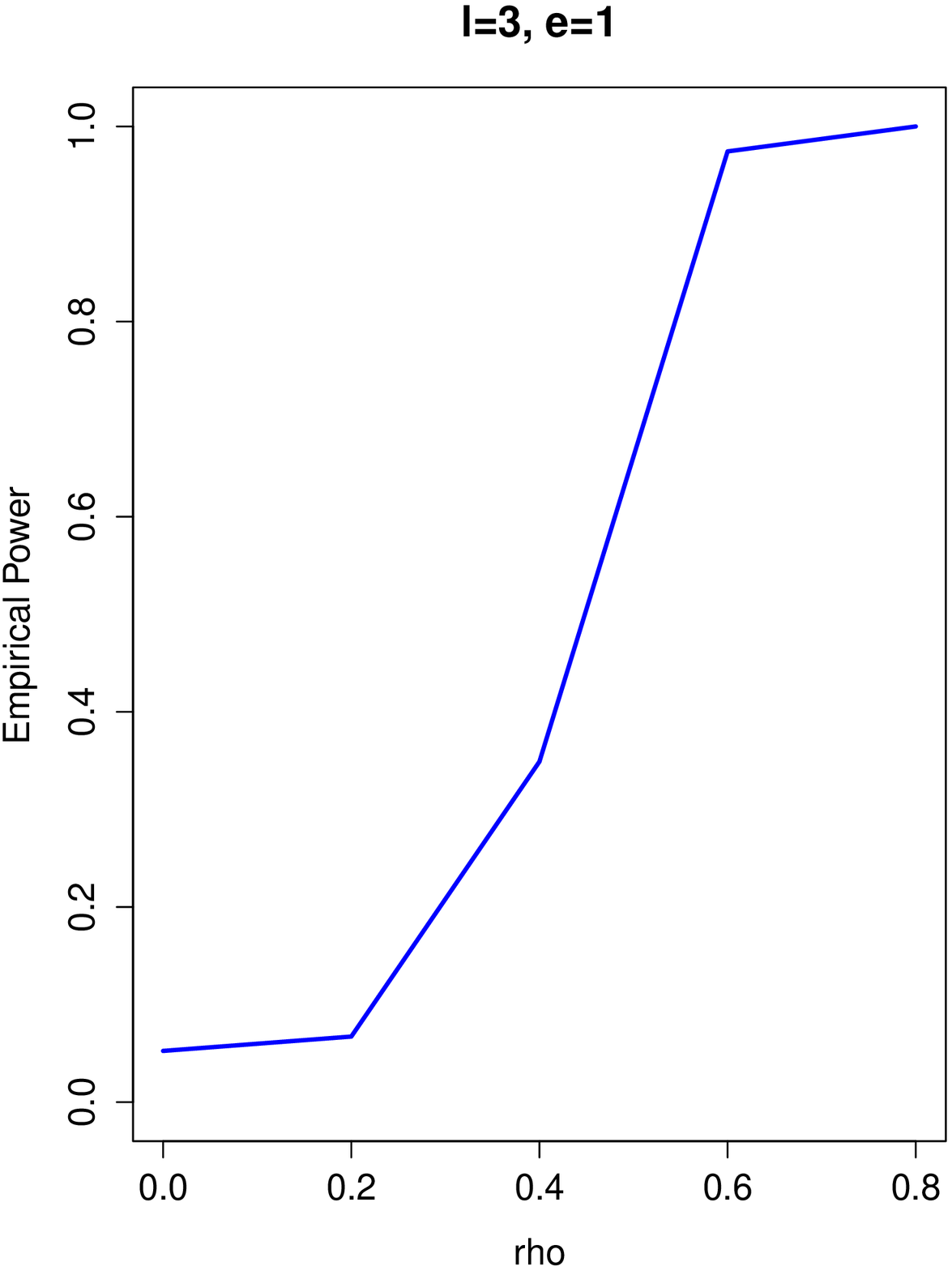}
\includegraphics[height=6cm,width=4cm]{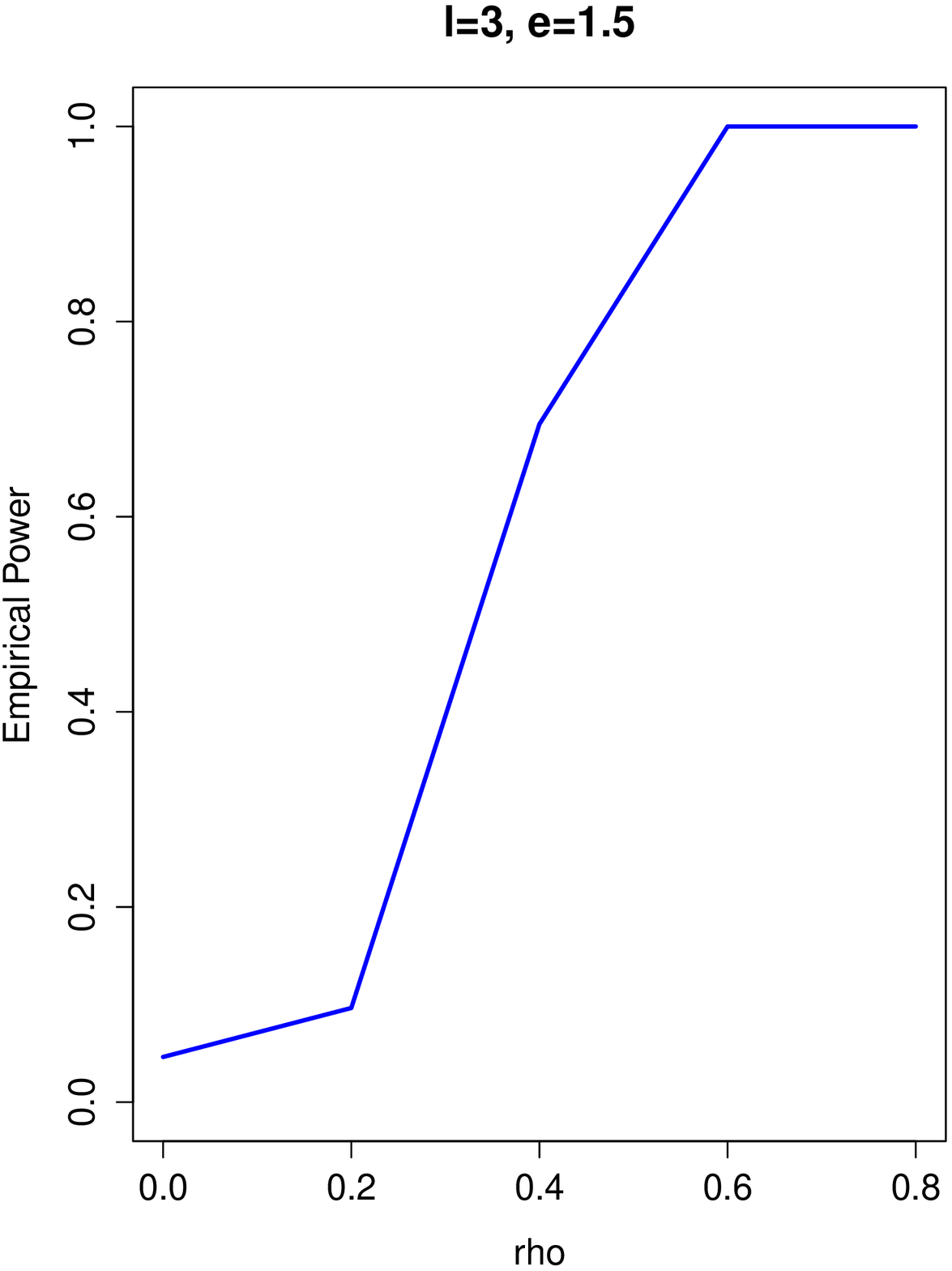}
\ \end{center} \small{Note: $L_x = L_y = m = 1$ in our test. Simulation is conducted with the test level $\alpha= 5\%$, and 10,000
replications.}
\label{figurecase1}
\end{figure}

\begin{figure}[!htbp]
\caption{Test nonlinear Granger causality form $Y_t$ to $X_t$: Case 2}
\begin{center} \
\includegraphics[height=6cm,width=4cm]{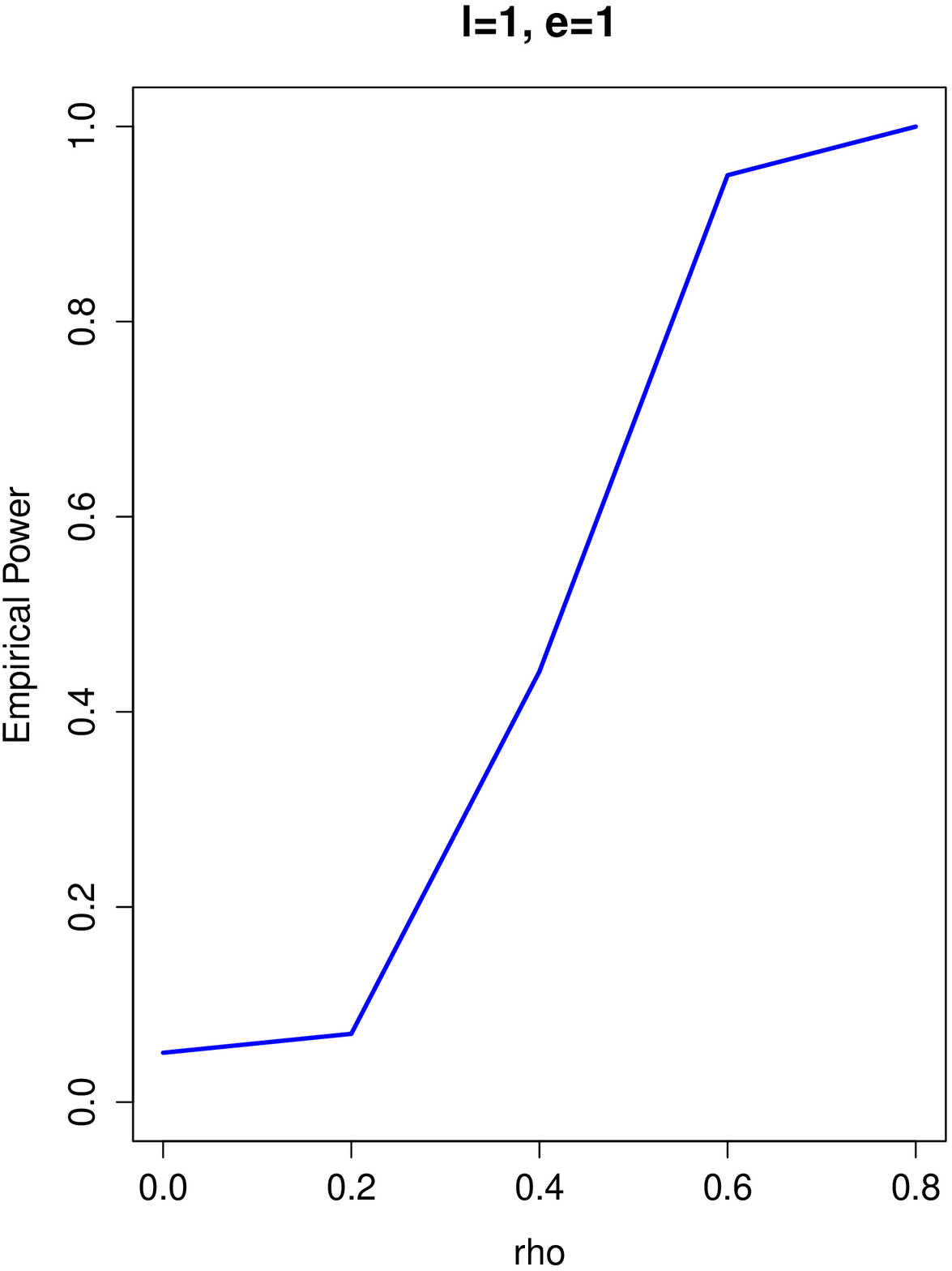}
\includegraphics[height=6cm,width=4cm]{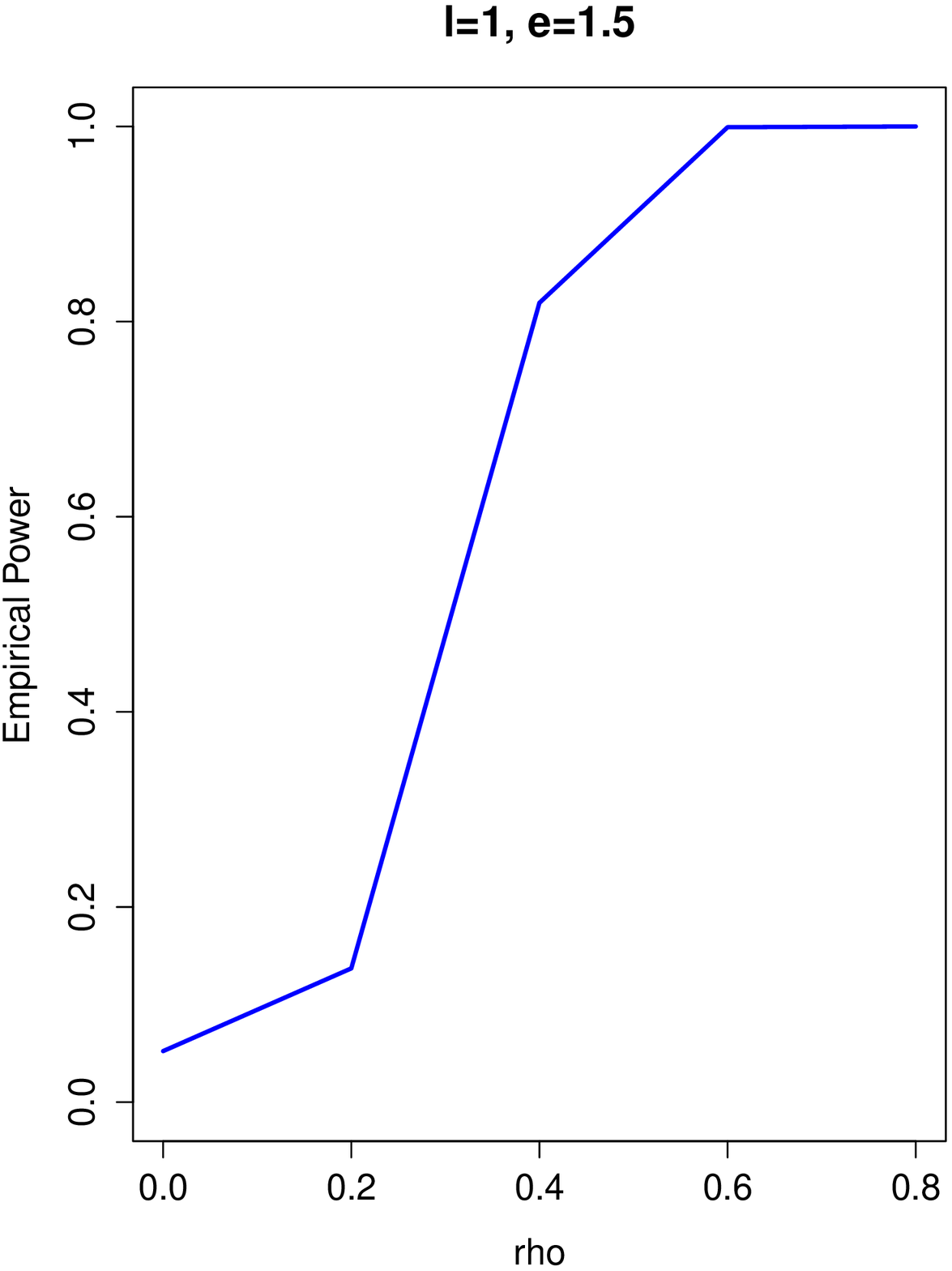} \\
\includegraphics[height=6cm,width=4cm]{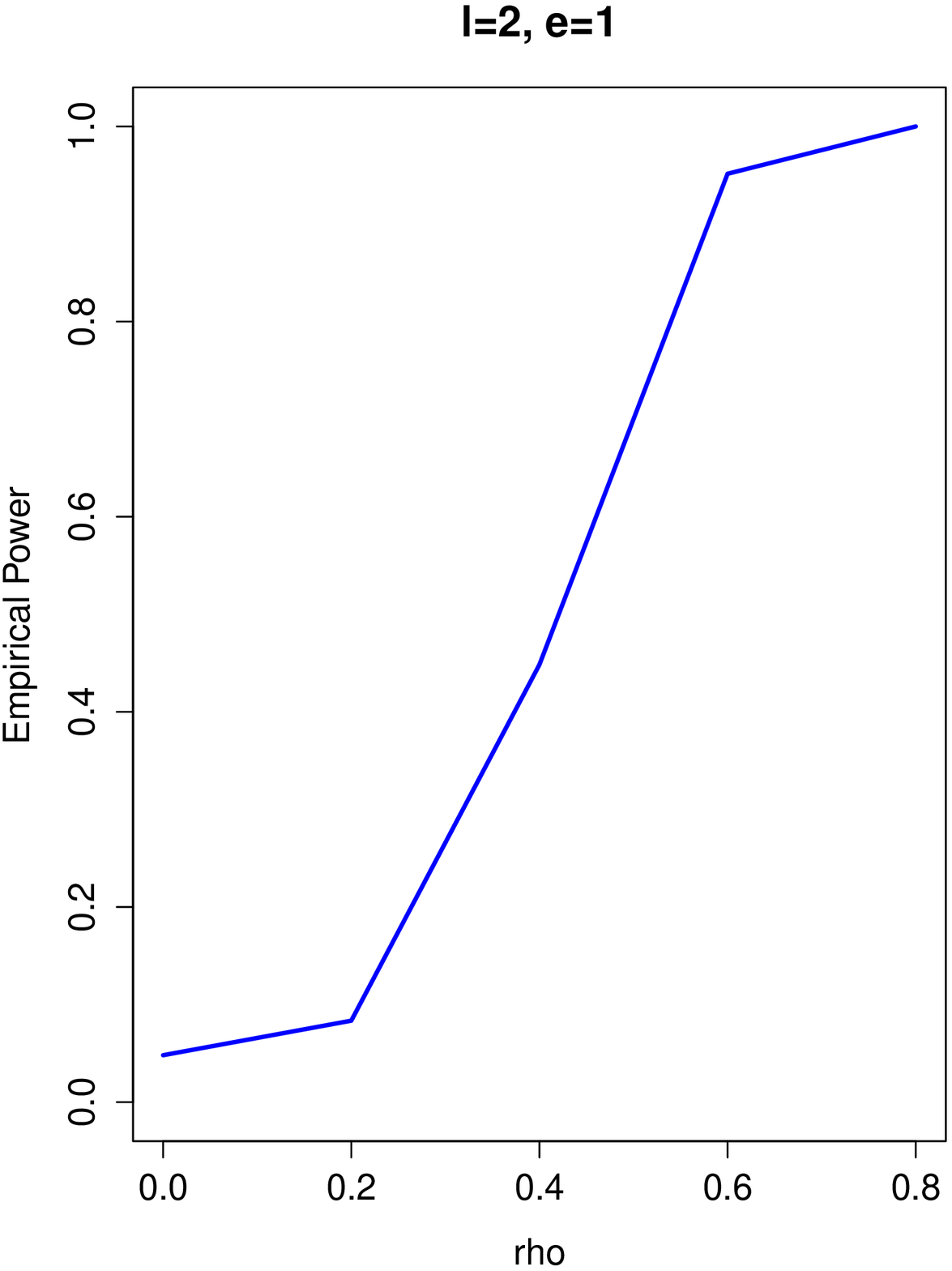}
\includegraphics[height=6cm,width=4cm]{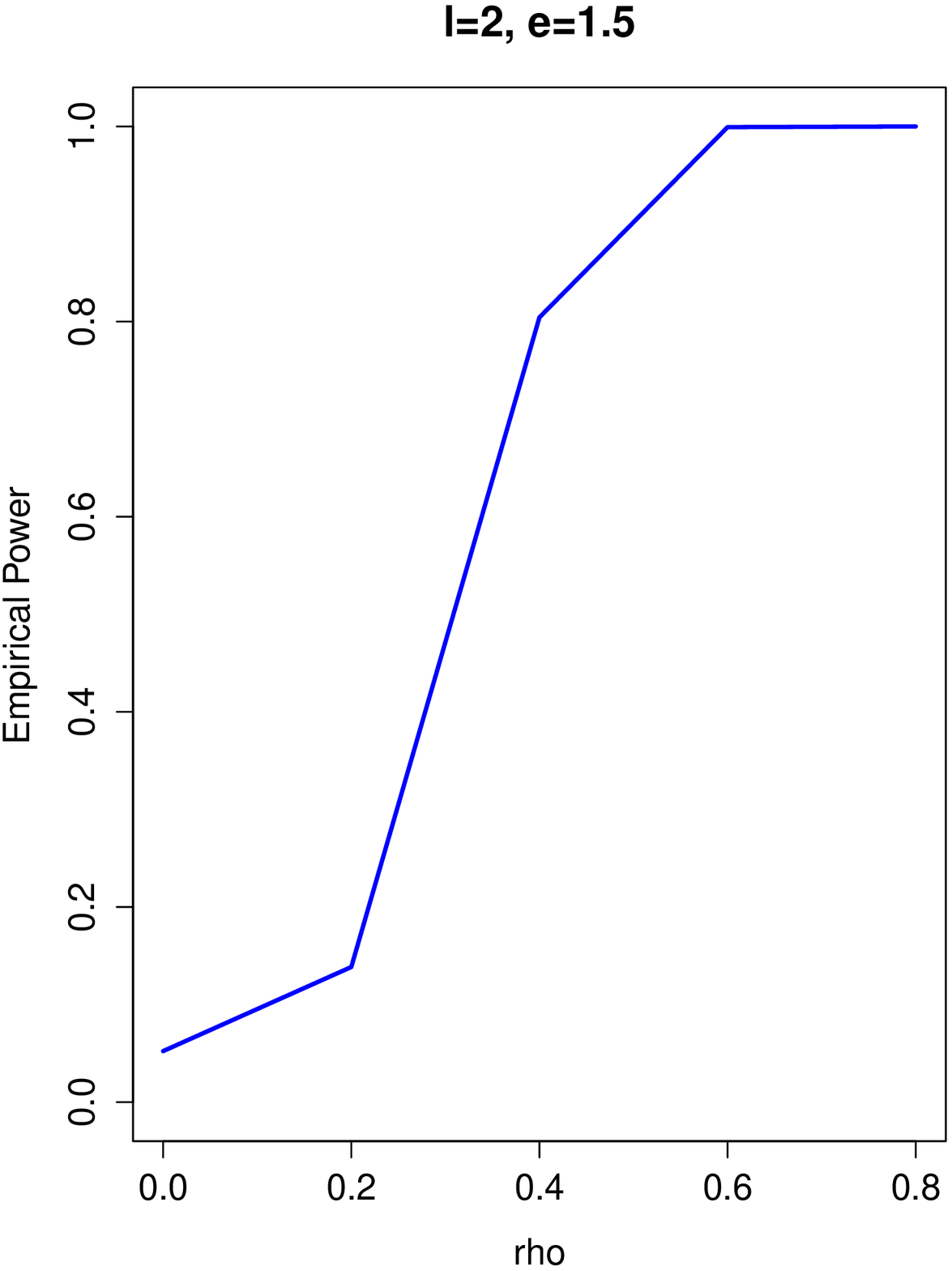}  \\
\includegraphics[height=6cm,width=4cm]{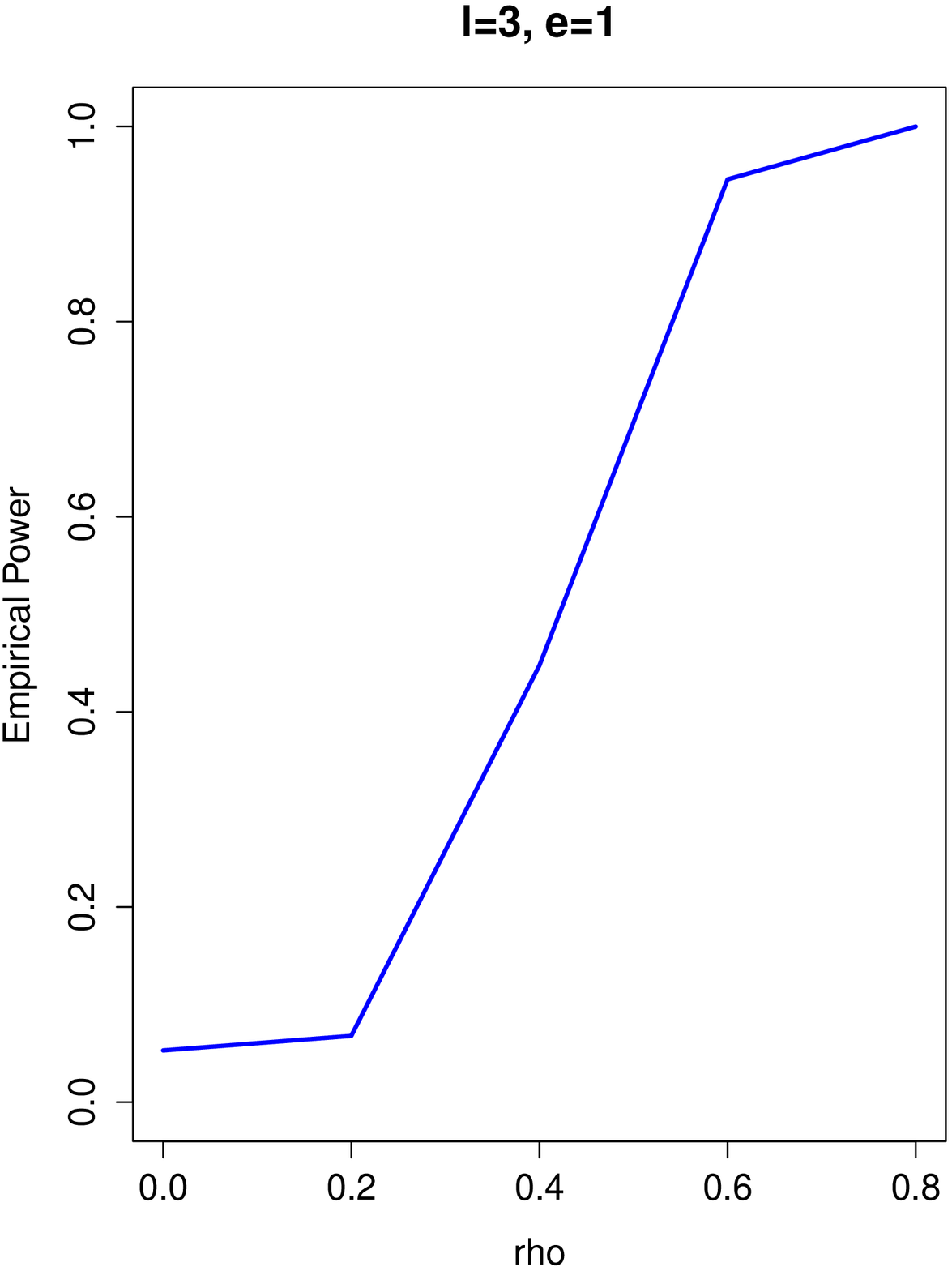}
\includegraphics[height=6cm,width=4cm]{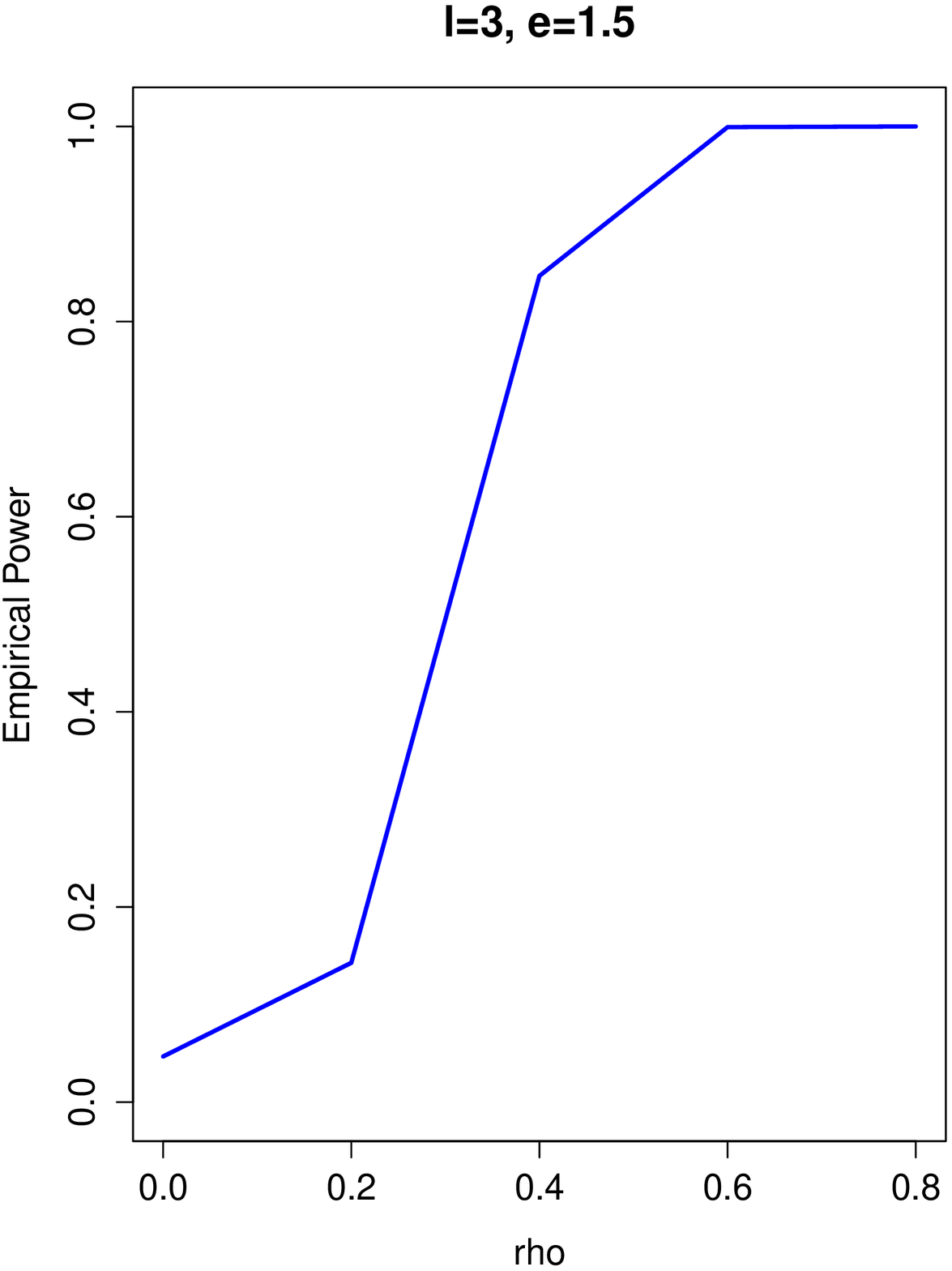}
\ \end{center} \small{Note: $L_x = L_y = m = 1$ in our test. Simulation is conducted with the test level $\alpha= 5\%$, and 10,000
replications.}
\label{figurecase2}
\end{figure}

In neither case is there nonlinear Granger causality from $Y_t$ to $X_t$ when $\rho= 0$, and causality strengthens when $\rho$
 increases. From the results displayed in Figure 1 and Figure 2, we conclude first that there is no
over-rejection problem in our new test. Second, our test possesses very appropriate power, as we see that empirical power sharply increases to 1 as $\rho$ increases. Further, we find that different settings
of $e$ may influence detection, and we suggest that practitioners  choose a couple of different values.

\newpage
\vspace{.2in}
\section{Illustration}\label{conclusion}
In this section, we consider an application to daily trading volumes and prices data for the Standard and Poor's 500 index over the period from January 2001 to October 2016. We denote the price and trading volume at $t$ as $P_t$ and $V_t$. Following Hiemstra and Jones (1994), the daily stock returns and the percentage change in trading volume are
expressed as $100\cdot\ln({P_t}/{P_{t-1}})$ and $100\cdot\ln({V_t}/{V_{t-1}})$, respectively.
We apply our new test to the daily stock returns and percentage changes in trading volume directly, since we do not emphasize
nonlinear Granger causality over linear Granger causality. 
In our test, we let $L_x = L_y = m = 1$. We consider three different settings for $l$ and two different settings for $e$, which are the same as those used in our simulation study.

\bigskip
The results presented in Table 2 show that Granger causality exists
in both directions between stock returns and volume changes.
In the first panel, when $l=2$ and $3$ for both $e=1$ and $1.5$, our tests reject the null hypothesis, which is evidence of returns affecting future volume changes.
For the other causal direction, the evidence that volume changes affect returns is provided in the second panel.
As we can see, when $l=2$ our test rejects the null hypothesis at the 0.05 level for both $e=1$ and $1.5$;
moreover, our test rejects the null hypothesis at level 0.01 for $l=3$ with both settings of $e$. We should also note
that the test results are not significant when $l=1$ in the first panel or when $l=2$ in the second panel. We suggest
that practitioners may need to choose several settings of $l$.

\bigskip
Our findings are the same as those of Hiemstra and Jones (1994) and Diks and Panchenko (2006). We are not surprise that stock returns and volume changes cause each other, and this seems to be common sense to investors.
Though we admire the work of Hiemstra and Jones (1994) on the causal relationship between stock returns and volume changes,
we still suggest that practitioners reconsider conclusions that are obtained by the HJ test.

\begin{table}[!htbp]\scriptsize
\begin{center}
\caption{Tests of Granger causality between daily returns and volume changes in the S\&P 500 index, January 2001 to October 2016.}  \label{tableIllus} \vskip 0.1in
\begin{tabular}{|c|ccccccccc|}
\hline \multicolumn{1}{|c|}{$H_0: $ \ No Granger causality} &\multicolumn{1}{c}{}
&\multicolumn{3}{c}{$e=1$} &\multicolumn{1}{c}{}
&\multicolumn{3}{c}{$e=1.5$} &\multicolumn{1}{c|}{}\\
\cline{3-5}\cline{7-9}
from returns to volume changes && $l=1$ &$l=2$&$l=3$&&$l=1$&$l=2$& $l=3$& \\
\hline
$p$-value&&0.1442&$0.0184^{\ast\ast}$&$0.0287^{\ast\ast}$&&0.1833 &$0.0094^{\ast\ast\ast}$& $0.0143^{\ast\ast}$& \\
\hline
\hline \multicolumn{1}{|c|}{$H_0: $ \ No Granger causality} &\multicolumn{1}{c}{}
&\multicolumn{3}{c}{$e=1$} &\multicolumn{1}{c}{}
&\multicolumn{3}{c}{$e=1.5$} &\multicolumn{1}{c|}{}\\
\cline{3-5}\cline{7-9}
from volume changes to returns  && $l=1$ &$l=2$&$l=3$&&$l=1$&$l=2$& $l=3$& \\
\hline
$p$-value&&$0.0205^{\ast\ast}$&0.1221&$0.0091^{\ast\ast\ast}$&&$0.0402^{\ast\ast}$&$0.084^{\ast}$&$0.0003^{\ast\ast\ast}$& \\
\hline
\end{tabular}\\
\begin{flushleft}
{\small Note: In our test, we chose $L_x=L_y=m=1$. ``$\ast$'', ``$\ast\ast$'' and ``$\ast\ast\ast$'' indicate significance at the 0.1, 0.05 and 0.01 levels, respectively.}
\end{flushleft}
\end{center}
\end{table}

\vspace{.2in}\section{Concluding Remarks}\label{conclusion}
\bigskip
Hiemstra and Jones's pioneering work on a modified version of
the nonlinear causality test in Baek and Brock (1992) is a significant breakthrough in the history of causal inference, since Baek and Brock (1992)
assume that the time series to which the test is applied are mutually independent and individually independent and identically
distributed. Most importantly, Hiemstra and Jones illustrated a promising nonparametric approach to causality testing to uncover significant nonlinearities in the dynamic interrelationships between economic variables.

\bigskip
Hiemstra and Jones (1994) has encouraged thousands of works, both theoretical and practical, over the last two decades.
In this paper, we reveal the underlying reasons for the questionable performance of the HJ test.
We find that Hiemstra and Jones (1994) propose inconsistent estimators of $C_1 \big({m}+{L_x},{L_y},e \big)$, $C_2 \big({L_x},{L_y},e \big)$, $C_3 \big (m+{L_x},e \big )$ and $C_4 \big ({L_x},e \big )$. Further, the $U$-statistics used to prove the asymptotic property of their test statistics are not valid, since there is no $U$-statistic.

\bigskip
By re-estimating the probabilities in the definitions, we propose a new, straightforward
test statistic to test the same null hypothesis tested in Hiemstra and Jones (1994).
The simulations show that our new test possesses acceptable power and, most importantly, that there is no
over-rejection problem.

\bigskip
We should still note some of the limitations of the HJ test. For example, Diks and Panchenko (2006) highlight a need for substitutions for the relationship tested in the Hiemstra-Jones test. We will look for a more
appropriate relationship to describe that $\{Y_t\}$ is not a Granger cause of $\{X_t\}$.

\section*{Appendix}

\setcounter{equation}{0}

\def\theequation{A.\arabic{equation}}

\bigskip
\noindent

\bigskip \noindent
{\large \bf A1: Central Limit Theorems for strong mixing stationary sequence}

\bigskip
\noindent
$\{\left(Z_t, \mathcal{F}_t\right), -\infty < t < \infty\}$ is a stochastic process defined on the probability space $(\Omega,\mathcal{F},P)$.
The history and the future of $Z_t$  are $\sigma$-algebras $\mathfrak{M}_{t}^\infty = \{\mathcal{F}_{s},s > t\}$ and $\sigma$-algebras
$\mathfrak{M}_{-\infty}^{t} = \{\mathcal{F}_{s},s < t\}$ respectively.

Let \{$\left(Z_i, \mathcal{F}_i\right)$\} be a stationary sequence with $E(Z_i)=0$, $E({Z_i}^2)<0$,and set $S_n^m=\sum\limits_{i=m}^{n+m}Z_i$,${\sigma_n}^2=Var(S_n^m)$.We shall say that the sequence satisfies the central limit theorem if
\begin{eqnarray*}
\lim_{n\to\infty}P\{\frac{S_n^m}{\sigma_n}<z\}=(2\pi)^{-\frac{1}{2}}\int_{-\infty}^{z}{e^{-\frac{1}{2}u^2}du}=\Phi(z) \ .
\end{eqnarray*}

\bigskip \noindent
{\bf Definition A1:}{ A stationary process \{$Z_t$\} is said to be strongly mixing (completely regular) if
$\alpha(\tau)=\sup\limits_{A\in {\mathfrak{m}_{-\infty}^0}, {B \in \mathfrak{m}_\tau^\infty}}|P(AB)-P(A)P(B)|\to 0$
as $\tau \to \infty$ through positive values.}

\noindent \textbf{Lemma A1}: Let the stationary sequence \{$Z_i$\} satisfy the strong mixing condition with mixing coefficient $\alpha(n)$, and let $E|Z_i|^{2+\delta}<\infty$ for some $\delta>0$. If $\sum\limits_{n=1}^{\infty}{\alpha(n)}^{\delta/(2+\delta)}<\infty$, then $\sigma^2=E({Z_0}^2)+2\sum\limits_{j=1}^{\infty}E(Z_0Z_j)<\infty$, and if $\sigma\neq0$, then $\lim\limits_{n\to\infty}P\{\sigma^{-1}n^{-\frac{1}{2}}\sum\limits_{i=1}^{n}Z_i<z\}=\Phi(z)$.

Readers can refer to Ibragimov (1971) for a proof and detailed discussion.

\bigskip
\noindent
 {\large \bf A2: Proof of Theorem \ref{BHWtest}}

\bigskip \noindent
Assume $\{x_1, x_2, \cdots  ,x_T\}$ and $\{y_1, y_2, \cdots  ,y_T\}$ are both strong mixing stationary sequences whose
 mixing coefficient satisfying the conditions in Lemma 1. Then the following four sequences
\begin{eqnarray*}
&& \{Z_{1t} = I \big(x_{t-{L_x}}^{m+{L_x}},x_{t+l-{L_x}}^{m+{L_x}},e  \big) \cdot
I \big(y_{t-L_y}^{L_y},y_{t+l-L_y}^{L_y},e \big) - {C}_1 \big (m+{L_x},L_y,e;l \big )\} \, , \\
&& \{Z_{2t} = I \big(x_{t-{L_x}}^{{L_x}},x_{t+l-{L_x}}^{{L_x}},e \big)
\cdot I \big(y_{t-L_y}^{L_y},y_{t+l-L_y}^{L_y},e \big) - {C}_2  \big ({L_x},L_y,e;l \big )\} \, ,  \\
&& \{Z_{3t} = I \big(x_{t-{L_x}}^{m+{L_x}}, x_{t+l-{L_x}}^{m+{L_x}},e \big) - {C}_3 \big (m+{L_x},e; l \big )\} \, ,  \\
&& \{Z_{4t} = I \big(x_{t-{L_x}}^{{L_x}},x_{t+l-{L_x}}^{{L_x}},e \big) - {C}_4 \big ({L_x},e;l \big )\} \, , t = L_{xy}+1, \cdots, T-l-L_{xy}-m+1,
\end{eqnarray*}
satisfy the conditions of Lemma 1. So \{$Z_{1t}$\}, \{$Z_{2t}$\}, \{$Z_{3t}$\} and \{$Z_{4t}$\} satisfy the central limit theorem.

Further, for any $a_1, a_2, a_3$ and $a_4$, sequence \{$Z_t = a_1 Z_{1t} + a_2 Z_{2t} + a_3 Z_{3t} + a_4 Z_{4t}, t = L_{xy}, \cdots, T-l-L_{xy}-m+1$\}
 also satisfies the conditions of Lemma 1 which implying that

\begin{eqnarray*}
 \sqrt{n}\left[
    \begin{array}{c}
\hat{C}_1  \big (m+{L_x},L_y,e;l \big )-{C}_1  \big (m+{L_x},L_y,e;l \big )\\
\hat{C}_2  \big ({L_x},L_y,e;l \big )-{C}_2  \big ({L_x},L_y,e;l \big )\\
\hat{C}_3  \big (m+{L_x},e;l \big )-{C}_3  \big (m+{L_x},e;l \big )\\
\hat{C}_4  \big (L_x,e;l \big )-{C}_4 \big ({L_x},e;l \big )
    \end{array} \right]
\overset{d}{\longrightarrow} N(0,\mathbf{\Sigma}),
\end{eqnarray*}
where $\mathbf{\Sigma}$ is a $4\times4$ symmetric matrix. Denote
\begin{eqnarray*}
&& h_{1}(L_x,L_y,m,l,k) = I(x^{L_x +m}_{L_{xy} +1+k-L_x}, x^{L_x +m}_{L_{xy}+1+k+l -L_x}, e) \, , \\
&& h_{2}(L_x,L_y,l,k) = I(y^{L_y}_{L_{xy}+1+k-L_y}, y^{L_y}_{L_{xy}+1+k+l -L_y}, e) \ ,
\end{eqnarray*}
we have
\begin{eqnarray*}
&&\mathbf{\Sigma}_{11} = E \left[\big(h_{1}(L_x,L_y,m,l,0)h_{2}(L_x,L_y,l,0) - {C}_1 (m+{L_x},L_y,e;l )\big)^2 \right]\\
&&+  \sum\limits_{k=1}^{n-1} 2(1-\frac{k}{n})E \Big[\big(h_{1}(L_x,L_y,m,l,0)h_{2}(L_x,L_y,l,0) - {C}_1 (m+{L_x},L_y,e;l )\big)\\
&&\ \ \ \ \ \ \ \ \ \ \ \ \ \ \ \ \ \ \ \ \ \ \     \big(h_{1}(L_x,L_y,m,l,k)h_{2}(L_x,L_y,l,k) - {C}_1 (m+{L_x},L_y,e;l )\big)\Big] \ ,
\\
&&\mathbf{\Sigma}_{12} = E \Big[\big(h_{1}(L_x,L_y,m,l,0)h_{2}(L_x,L_y,l,0) - {C}_1 (m+{L_x},L_y,e;l )\big) \\
&& \ \ \ \ \ \ \ \ \ \ \ \ \ \big(h_{1}(L_x,L_y,0,l,0)h_{2}(L_x,L_y,l,0) - {C}_2 ({L_x},L_y,e;l )\big) \Big]\\
&&+  \sum\limits_{k=1}^{n-1} (1-\frac{k}{n})E \Big[\big(h_{1}(L_x,L_y,m,l,0)h_{2}(L_x,L_y,l,0) - {C}_1 (m+{L_x},L_y,e;l )\big)\\
&&\ \ \ \ \ \ \ \ \ \ \ \ \ \ \ \ \ \ \ \ \ \  \big(h_{1}(L_x,L_y,0,l,k)h_{2}(L_x,L_y,l,k) - {C}_2 ({L_x},L_y,e;l )\big)\Big]\\
&&+  \sum\limits_{k=1}^{n-1} (1-\frac{k}{n})E \Big[\big(h_{1}(L_x,L_y,m,l,k)h_{2}(L_x,L_y,l,k) - {C}_1 (m+{L_x},L_y,e;l )\big)\\
&&\ \ \ \ \ \ \ \ \ \ \ \ \ \ \ \ \ \ \ \ \ \  \big(h_{1}(L_x,L_y,0,l,0)h_{2}(L_x,L_y,l,0) - {C}_2 ({L_x},L_y,e;l )\big)\Big] \ ,
\\
&&\mathbf{\Sigma}_{13} = E \Big[\big(h_{1}(L_x,L_y,m,l,0)h_{2}(L_x,L_y,l,0) - {C}_1 (m+{L_x},L_y,e;l )\big) \\
&& \ \ \ \ \ \ \ \ \ \ \ \ \ \big(h_{1}(L_x,L_y,m,l,0) - {C}_3 ({m+L_x},e;l )\big) \Big]\\
&&+  \sum\limits_{k=1}^{n-1} (1-\frac{k}{n})E \Big[\big(h_{1}(L_x,L_y,m,l,0)h_{2}(L_x,L_y,l,0) - {C}_1 (m+{L_x},L_y,e;l )\big)\\
&&\ \ \ \ \ \ \ \ \ \ \ \ \ \ \ \ \ \ \ \ \ \  \big(h_{1}(L_x,L_y,m,l,k) - {C}_3 ({m+L_x},e;l )\big)\Big]\\
&&+  \sum\limits_{k=1}^{n-1} (1-\frac{k}{n})E \Big[\big(h_{1}(L_x,L_y,m,l,k)h_{2}(L_x,L_y,l,k) - {C}_1 (m+{L_x},L_y,e;l )\big)\\
&&\ \ \ \ \ \ \ \ \ \ \ \ \ \ \ \ \ \ \ \ \ \  \big(h_{1}(L_x,L_y,m,l,0) - {C}_3 ({m+L_x},e;l )\big)\Big] \ ,
\\
&&\mathbf{\Sigma}_{14} = E \Big[\big(h_{1}(L_x,L_y,m,l,0)h_{2}(L_x,L_y,l,0) - {C}_1 (m+{L_x},L_y,e;l )\big) \\
&& \ \ \ \ \ \ \ \ \ \ \ \ \ \big(h_{1}(L_x,L_y,0,l,0) - {C}_4 ({L_x},e;l )\big) \Big]\\
&&+  \sum\limits_{k=1}^{n-1} (1-\frac{k}{n})E \Big[\big(h_{1}(L_x,L_y,m,l,0)h_{2}(L_x,L_y,l,0) - {C}_1 (m+{L_x},L_y,e;l )\big)\\
&&\ \ \ \ \ \ \ \ \ \ \ \ \ \ \ \ \ \ \ \ \ \  \big(h_{1}(L_x,L_y,0,l,k) - {C}_4 ({L_x},e;l )\big)\Big]\\
&&+  \sum\limits_{k=1}^{n-1} (1-\frac{k}{n})E \Big[\big(h_{1}(L_x,L_y,m,l,k)h_{2}(L_x,L_y,l,k) - {C}_1 (m+{L_x},L_y,e;l )\big)\\
&&\ \ \ \ \ \ \ \ \ \ \ \ \ \ \ \ \ \ \ \ \ \  \big(h_{1}(L_x,L_y,0,l,0) - {C}_4 ({L_x},e;l )\big)\Big] \ ,
\end{eqnarray*}
\begin{eqnarray*}
&&\mathbf{\Sigma}_{22} = E \left[\big(h_{1}(L_x,L_y,0,l,0)h_{2}(L_x,L_y,l,0) - {C}_2 ({L_x},L_y,e;l )\big)^2 \right]\\
&&+  \sum\limits_{k=1}^{n-1} 2(1-\frac{k}{n})E \Big[\big(h_{1}(L_x,L_y,0,l,0)h_{2}(L_x,L_y,l,0) - {C}_2 ({L_x},L_y,e;l )\big)\\
&&\ \ \ \ \ \ \ \ \ \ \ \ \ \ \ \ \ \ \ \ \ \ \     \big(h_{1}(L_x,L_y,0,l,k)h_{2}(L_x,L_y,l,k) - {C}_2 ({L_x},L_y,e;l )\big)\Big] \ ,
\\
&&\mathbf{\Sigma}_{23} = E \Big[\big(h_{1}(L_x,L_y,0,l,0)h_{2}(L_x,L_y,l,0) - {C}_2 ({L_x},L_y,e;l )\big) \\
&& \ \ \ \ \ \ \ \ \ \ \ \ \ \big(h_{1}(L_x,L_y,m,l,0) - {C}_3 (m+{L_x},e;l )\big) \Big]\\
&&+  \sum\limits_{k=1}^{n-1} (1-\frac{k}{n})E \Big[\big(h_{1}(L_x,L_y,0,l,0)h_{2}(L_x,L_y,l,0) - {C}_2 ({L_x},L_y,e;l )\big)\\
&&\ \ \ \ \ \ \ \ \ \ \ \ \ \ \ \ \ \ \ \ \ \  \big(h_{1}(L_x,L_y,m,l,k) - {C}_3 (m+{L_x},e;l )\big)\Big]\\
&&+  \sum\limits_{k=1}^{n-1} (1-\frac{k}{n})E \Big[\big(h_{1}(L_x,L_y,0,l,k)h_{2}(L_x,L_y,l,k) - {C}_2 ({L_x},L_y,e;l )\big)\\
&&\ \ \ \ \ \ \ \ \ \ \ \ \ \ \ \ \ \ \ \ \ \  \big(h_{1}(L_x,L_y,m,l,0) - {C}_3 (m+{L_x},e;l )\big)\Big] \ ,
\\
&&\mathbf{\Sigma}_{24} = E \Big[\big(h_{1}(L_x,L_y,0,l,0)h_{2}(L_x,L_y,l,0) - {C}_2 ({L_x},L_y,e;l )\big) \\
&& \ \ \ \ \ \ \ \ \ \ \ \ \ \big(h_{1}(L_x,L_y,0,l,0) - {C}_4 ({L_x},e;l )\big) \Big]\\
&&+  \sum\limits_{k=1}^{n-1} (1-\frac{k}{n})E \Big[\big(h_{1}(L_x,L_y,0,l,0)h_{2}(L_x,L_y,l,0) - {C}_2 ({L_x},L_y,e;l )\big)\\
&&\ \ \ \ \ \ \ \ \ \ \ \ \ \ \ \ \ \ \ \ \ \  \big(h_{1}(L_x,L_y,0,l,k) - {C}_4 ({L_x},e;l )\big)\Big]\\
&&+  \sum\limits_{k=1}^{n-1} (1-\frac{k}{n})E \Big[\big(h_{1}(L_x,L_y,0,l,k)h_{2}(L_x,L_y,l,k) - {C}_2 ({L_x},L_y,e;l )\big)\\
&&\ \ \ \ \ \ \ \ \ \ \ \ \ \ \ \ \ \ \ \ \ \  \big(h_{1}(L_x,L_y,0,l,0) - {C}_4 ({L_x},e;l )\big)\Big] \ ,
\\
&&\mathbf{\Sigma}_{33} = E \left[\big(h_{1}(L_x,L_y,m,l,0) - {C}_3 (m+{L_x},e;l )\big)^2 \right]\\
&&+  \sum\limits_{k=1}^{n-1} 2(1-\frac{k}{n})E \Big[\big(h_{1}(L_x,L_y,m,l,0) - {C}_3 (m+{L_x},e;l )\big)\\
&&\ \ \ \ \ \ \ \ \ \ \ \ \ \ \ \ \ \ \ \ \ \ \     \big(h_{1}(L_x,L_y,m,l,k) - {C}_3 (m+{L_x},e;l )\big)\Big] \ ,
\end{eqnarray*}
\begin{eqnarray*}
&&\mathbf{\Sigma}_{34} = E \Big[\big(h_{1}(L_x,L_y,m,l,0) - {C}_3 (m+{L_x},e;l )\big) \\
&& \ \ \ \ \ \ \ \ \ \ \ \ \ \big(h_{1}(L_x,L_y,0,l,0) - {C}_4 ({L_x},e;l )\big) \Big]\\
&&+  \sum\limits_{k=1}^{n-1} (1-\frac{k}{n})E \Big[\big(h_{1}(L_x,L_y,m,l,0) - {C}_3 (m+{L_x},e;l )\big)\\
&&\ \ \ \ \ \ \ \ \ \ \ \ \ \ \ \ \ \ \ \ \ \  \big(h_{1}(L_x,L_y,0,l,k) - {C}_4 ({L_x},e;l )\big)\Big]\\
&&+  \sum\limits_{k=1}^{n-1} (1-\frac{k}{n})E \Big[\big(h_{1}(L_x,L_y,m,l,k) - {C}_3 (m+{L_x},e;l )\big)\\
&&\ \ \ \ \ \ \ \ \ \ \ \ \ \ \ \ \ \ \ \ \ \  \big(h_{1}(L_x,L_y,0,l,0) - {C}_4 ({L_x},e;l )\big)\Big] \ ,
\\
&&\mathbf{\Sigma}_{44} = E \left[\big(h_{1}(L_x,L_y,0,l,0) - {C}_4 ({L_x},e;l )\big)^2 \right]\\
&&+  \sum\limits_{k=1}^{n-1} 2(1-\frac{k}{n})E \Big[\big(h_{1}(L_x,L_y,0,l,0) - {C}_4 ({L_x},e;l )\big)\\
&&\ \ \ \ \ \ \ \ \ \ \ \ \ \ \ \ \ \ \ \ \ \ \     \big(h_{1}(L_x,L_y,0,l,k) - {C}_4 ({L_x},e;l )\big)\Big] \ .
\end{eqnarray*}

Under the null hypothesis, applying the delta method (Serfling, 1980), we have
\begin{align*}
\sqrt{n}\left(\frac{\hat{C}_1 \big (m+{L_x},L_y,e,l \big )}{\hat{C}_2 \big
({L_x},L_y,e,l \big )}-\frac{\hat{C}_3 \big (m+{L_x},e,l \big )}{\hat{C}_4 \big
({L_x},e,l \big )}\right) \overset{d}{\longrightarrow} N \big(0, \sigma^2
(m,{L_x},L_y,e,l) \big) \, ,
\end{align*}
where
$\sigma^2(m,{L_x},L_y,e,l) = \nabla^{\prime}{\mathbf{\Sigma}}\nabla$, in which
\begin{align*}
\nabla &= \left( \frac 1 { C_2 \big({L_x},{L_y},e,l \big)} \, , \, - \frac
{ C_1 \big(m+{L_x},{L_y},e,l \big)}{C_2^2 \big({L_x},{L_y},e,l
\big)}  \, , \, - \frac 1 {C_4 \big({L_x},e,l \big)}  \, , \, \frac
{ C_3 \big({M_x}+{L_x},e,l \big)}{C_4^2 \big({L_x},e,l \big)} \right)^{\prime} \ .
\end{align*}

An consistent estimator $\hat{\sigma}^2(m,{L_x},L_y,e,l)$ of the asymptotic variance
can be got by replacing all the parts in the sandwich $\nabla^{\prime}{\mathbf{\Sigma}}\nabla$ by their empirical estimates.

This completes the proof of the theorem.
\text{ } \hfill $\Box $

\end{document}